\documentclass[compsoc,conference,a4paper,10pt,times]{IEEEtran}
\IEEEoverridecommandlockouts

\usepackage{ifthen}

\newboolean{xl} 
\setboolean{xl}{false} 

\newboolean{ieee}
\setboolean{ieee}{true} 
\newboolean{lncs}
\setboolean{lncs}{false} 
\newboolean{acm}
\setboolean{acm}{false} 
\newboolean{usenix}
\setboolean{usenix}{false} 

\ifthenelse{\boolean{usenix}}{
  \usepackage{usenix-2020-09}}{}%
\usepackage{mathtools}
\usepackage{epsfig,graphicx}
\ifthenelse{\boolean{acm}}{
  \usepackage{amsmath, amsfonts}
}{
  \usepackage{amsmath,amssymb,amsfonts}
}
\usepackage[font=itshape]{quoting}
\usepackage{svg}
\usepackage{underscore}
\usepackage[english]{babel}
\usepackage{calrsfs}
\usepackage[hyphens]{url}
\usepackage{gensymb}
\usepackage{multirow,colortbl}
\usepackage{adjustbox,lscape}
\usepackage{tikz}
\usetikzlibrary{positioning}
\usepackage{colortbl}
\usepackage{paralist}
\usepackage{algorithm,algorithmic}
\usepackage{multicol}
\usepackage{extarrows}
\usepackage{tabularx}
\usepackage{hhline}
\usepackage{wrapfig}
\usepackage{flushend}
\ifthenelse{\boolean{ieee}}{
  \usepackage[ breaklinks, colorlinks=true,	linkcolor=black]{hyperref}
  \usepackage{bmpsize}
  \usepackage{xcolor}
  \usepackage{lipsum}
  \usepackage{amsthm}
  }{}

\ifthenelse{\boolean{acm}}{}{
  \usepackage{verbatimbox}
}
\usepackage{graphicx}
\usepackage{subcaption}
\usepackage{url}
\usepackage{array}
\newcolumntype{L}[1]{>{\raggedright\let\newline\\\arraybackslash\hspace{0pt}}m{#1}}
\newcolumntype{C}[1]{>{\centering\let\newline\\\arraybackslash\hspace{0pt}}m{#1}}
\newcolumntype{R}[1]{>{\raggedleft\let\newline\\\arraybackslash\hspace{0pt}}m{#1}}

\ifthenelse{\boolean{ieee}}{%

}{}

\author{Daniel Migault}

\newcommand{\co}{$CO_2$}
\newcommand{\coeq}{\texttt{CO2eq}}
\newcommand{\thiscoeq}{\texttt{CO2eq-v0.0.1}}

\begin{document}

\title{\coeq{}: Estimating Meetings' Air Flight \co{} Equivalent Emissions\\ An Illustrative Example with IETF meetings }
\maketitle

\begin{abstract}
These notes describe \coeq{}~\cite{co2eq} a tool that estimates \co{} equivalent emissions associated with air traffic and applies it to the Internet Engineering Task Force (IETF), an international standard developing organization that meets 3 times a year.    
\coeq{} estimates that the participation to IETF meetings (by a single participant) generates as much \co{} equivalent as the \co{} emissions per capita of European countries generating their energy using coal -- like Germany or Poland for example~\cite{co2eq.io}.
This suggests some radical changes should be considered by the IETF.

According to the conclusion of the $26^{th}$ Conference of the Parties (COP26) from the United Nations Secretary-General António Guterres; in 2021, the number of meetings should be limited to a maximum of one meeting per year. In addition, the incorporation of sustainability principles into the IETF's strategy, should include, for example, 
increasing the effort to enhance the experience of 'remote' participation as well as 
adhering to programs (such as for example the United Nations Global Compact~\cite{ungc} and the caring for climate initiative~\cite{c4c}) to align its strategy and report progress toward sustainability.

\end{abstract}



\vspace*{3mm}
Disclaimer: Opinions expressed are solely my own and do not express the views or opinions of my employer.
Ericsson has worked for a long time in research regarding ICT sustainability impact~\cite{eri-sustainability} and this work does not integrate yet these researches.
Specifically, we will align with methodological insights for assessing the environment effect induced by ICT services~\cite{10.1145/3401335.3401716,10.1145/3401335.3401711} in future version of the paper.

\section{Introduction}
\label{sec:intro}
On 2021 November 13 António Guterres, United Nations Secretary-General concludes the $26^{th}$ Conference of the Parties (COP26) in Glasgow with the following words~\cite{cop26}:

\begin{quoting}
\noindent Our fragile planet is hanging by a thread. \\ 
We are still knocking on the door of climate catastrophe.\\  
It is time to go into emergency mode — or our chance of reaching net zero will itself be zero.~(...) \\
Science tells us that the absolute priority must be rapid, deep and sustained emissions reductions in this decade.\\
Specifically — a 45\% cut by 2030 compared to 2010 levels.  \\
But the present set of Nationally Determined Contributions -- even if fully implemented -- will still increase emissions this decade on a pathway that will clearly lead us to well above 2 degrees by the end of the century compared to pre-industrial levels. (...) \\
COP 27 starts now.\\
 \begin{flushright}
    \small{---António Guterres, COP26}
  \end{flushright}
\end{quoting}

On the one hand, science urges us to reduce our emissions by 45\% to keep the temperature increase below +1.5 \degree C, 
we know global warming today has already reached +1.2 \degree C~\cite{climateclock}, 
we know the aviation sector is responsible of 2.5\% of the \co{} emissions and 3.5\% of the effective radiative forcing~\cite{owid-aviation} -- a more accurate measure of its contribution to global warming  that is referred as \co{} equivalent in these notes. 
However, we hardly see any strong commitment in reducing the frequency of international meetings which involve many international flights.

This paper describes \coeq{}~\cite{co2eq}, a tool that estimates the \co{} equivalent emissions generated by international conferences or meetings and applies it to the Internet Engineering Task Force (IETF)~\cite{ietf}, a standard developing organization meeting three times a year. 
Until March 2020 these meetings have always been held in person. During the COVID-19 pandemic, these meetings have been held entirely online, demonstrating online meetings are feasible and questioning the necessity of 3 'on-site' meetings in a normal situation.   
\coeq{} estimates the average amount of \co{} equivalent to an IETF attendee. 
\coeq{} shows \co{} emissions resulting from participating to the IETF is far from being negligible and attending 3 IETF meetings corresponds to the \co{} per capita of European countries producing energy based on coal -- such as Germany or Poland. 
Application of the 2015 Paris agreement would result in a 45\% cut of 2010 \co{} emission, which would mathematically limit the number of meetings to a single 'on-site' IETF meeting per year. 
To check the impact of the COVID-19 forcing remote meetings toward this initial goal, we considered scenarios envisioned for general aviation and applied them to air traffic associated with IETF participation. 
These simulations also  confirm the limitation to a single 'on-site' IETF meeting a year.
On the other hand, while coming by surprise and without any anticipation, 'remote' IETF meetings have shown to be working well and are very promising given that we are still in an adaptation mode, and the margin to improve the remote meeting experience remains huge.

In addition to \co{} emission, \coeq{} also estimates the number of flight connections per participant depending on the location. 
This data might be used for the selection of future meeting locations that limit the exposure of the attendees to viruses. Typically, it is expected the number of flight connections increases the exposure to viruses, and as such some places may be preferred than others. 
We also extend the estimation of \co{} equivalent to a more generic metric (such as the number of attendees) that could be used to evaluate growth and trends in IETF participation. 
Such considerations are mostly a starting point for a discussion and additional work and  further analysis are needed to come to draw conclusions.

The remaining of these notes is as follows: 
Section~\ref{sec:co2eq} details how \co{} equivalent emissions associated with air traffic are estimated. 
Section~\ref{sec:impl} details \coeq{}. 
Section~\ref{sec:ietf} comments and analyses \coeq{} outputs for IETF meetings and Section~\ref{sec:concl} concludes that potential actions to incorporate sustainability principles into the IETF’s strategy may include :
\begin{itemize}
\item[1] Limiting 'on-site' meetings to a maximum of 1 meeting a year.
\item[2] Increasing the effort to enhance the experience of 'remote' participation - in particular to address the issue of hallway meetings.
\item[3] Adhering to programs to ensure the IETF aligns its strategy and report progress toward sustainability -- such as for example the United Nations Global Compact~\cite{ungc} and the caring for climate initiative~\cite{c4c} .
\end{itemize}

\section{Flight \co{} Emission Estimation }
\label{sec:co2eq}
This section details the methodology used by \coeq{} to estimate flight \co{} equivalent emissions, and their respective implementation -- namely 'myclimate2018'~\cite{myclimate} and 'goclimate'~\cite{goclimate}.
Both start estimating the \co{} equivalent emissions associated to a flight, and then associate a proportion of it to each passenger. 

\subsection{Estimation of flight \co{} equivalent}

Most \co{} equivalent emissions during a flight ($E_{flight}$) is associated with the combustion of the fuel whose quantity depends on the category of aircraft, the flying distance as well as the different phases of the flight. 
Flights are usually decoupled into short haul and long haul aircraft with distinct consumption patterns. 
The different phases of a flight can be described as Landing and Take Off (LTO) or Climb Cruise Descent (CCD).  
The EMEP/EEA air pollutant emission inventory guidebook~\cite{eeaaviation1, eeaaviation2,eeaaviation3} provides for each type of aircraft the quantity of fuel burnt during LTO and CCD -- as well as the quantities of pollutant emitted on each phase.
To estimate the average consumed fuel per flying km -- across all acceptable aircraft --, ICAODATA provides the total distance flown by each aircraft (as well as the fuel consumption). 
This enables to derive a weighted average fuel consumption as a function of the flying distance $Fuel( d )$ in kg -- with $d$ the flying distance in km.
Note that $Fuel$ includes LTO.


The \co{} resulting from the combustion for 1 kg of fuel is $e_{CO_2}$ = 3.15 kg / kg of burnt fuel.
The impact of other non-\co{} pollutant affecting the earth radiative balance - such as nitrogen oxide ($NO_x$) are estimated through a Radiative Forcing Index ($RFI$) factor over the emissions of \co{} and~\cite{rfi} recommends to use $RFI$ = 2.
Note that the factor measures the effect of $NO_x$ and not the quantity. 
In fact $NO_x$ and \co{} have significant differences and in particular act on different time scales.
Outside of fuel combustion \co{}, one needs to consider the indirect source of emissions that is the \co{} emissions associated with fuel PreProcessing ($PP$) which is set to 0.54 kg / kg of burnt fuel. 

The flying distance between two airports considers the round shape of the earth -- using the Great Circle Distance -- as well as some extra Distance Correction ($DC$) due to inefficiency of the traffic control, weather conditions, and holding patterns. 

As a result \co{} equivalent emissions for a given flying distance $x = d + DC$ with $d$ the distance between the two airports can be expressed as:
\begin{equation}
E_{flight} ( x )  = Fuel(x) \times ( e_{CO_2}. RFI + PP )
\end{equation}

%

\subsection{\co{} equivalent per passenger}

The \co{} emissions per passenger $E_{flier}$ is estimated from $E_{flight}$ by considering the fraction of the load associated to the passenger, that is $1 - CARGO_{ld}$ with $CARGO_{ld}$ representing the cargo load.
This fraction of emissions is shared between the effective passengers weighted by the cabin class $W_{cabin}$ which is equivalent to occupying a certain number of economy seats.
The effective number of passengers is determined by the total capacity in term of seats $SEAT_{T}$ -- which depends on the aircraft type an can be found in ICAODATA -- multiplied by the load passenger factor $PSG_{ld}$ published by ICAO. 

As passenger and cargo are used to drive the demand for the construction of an airport or a plane, these are expressed on a per passenger basis. 
The aircraft life cycle is expressed as ($AIRCRAFT_{lc}$) is per passenger / per flying km and the infrastructures are modeled by a constant ($INFRA$)~\cite{lc}\cite{ecoinvent}.

As a result, the emissions per passenger are expressed as:
\begin{equation}
\begin{aligned}
E_{flier} (x)  = & E_{flight}(x) ( 1 - CARGO_{ld} ) \frac{ W_{cabin} }{ SEAT_{T} \times PSG_{ld} } \\
                 & + AIRCRAFT_{lc}. x + INFRA
\end{aligned}
\end{equation}

\subsection{ 'myclimate' versus 'goclimate'}

This section compares 'goclimate2019'~\cite{goclimate} as published on 2019-04-08 with 'myclimate2019'~\cite{myclimate} computation as published on 2019-08-13. 
\coeq{} implements 'myclimate2019', but relies on the service provided by GO Climate. 
As 'goclimate2019' references the latest version of 'myclimate' - in our case 'myclimate2019', we assume that the service synchronizes its principles with that latest version published by 'myclimate'. 

The 'myclimate' and 'goclimate' methodologies mostly differ in the estimation of the distance correction ($DC$) and the cargo load ($CARGO_{ld}$).  
'myclimate' considers a constant value for $DC$ = 95 km, while 'goclimate' respectively sets $DC$ to 50 km, 100 km and 125 km for flying distance respectively lower than 550 km, lower 5500 km and greater than 5500 km. 
In the case of the IETF where a significant number of flights are transcontinental $DC$ is increased between 5\% and 31\%. 
This is likely to increase the flying distance used by 'goclimate' and so the \co{} equivalent emissions. 
In addition, 'myclimate' estimates the cargo load ($CARGO_{ld}$) on a mass basis which is respectively 93\% for short haul and 74\% for long haul. 
On the other hand, 'goclimate' estimates the cargo load on a monetary basis to $CARGO_{ld}=95.1\%$. 
While 'goclimate' and 'myclimate' use ICAO as the source of information for the average number of seats and the passenger load ($PSG_{ld}$),  'goclimate' uses respectively ICAODATA~\cite{icaodata} 2012 and ICAO~\cite{icaoeco} 2012 while 'myclimate' respectively uses ICAODATA 2019 and ICAO 2018. 
More considerations may be needed to check if this presents an impact.

\section{\coeq{} Overview  }
\label{sec:impl}

\subsection{Overview}

Currently, \coeq~\cite{co2eq} estimates \co{} equivalent emissions associated with meetings. 
The \texttt{Meeting} class takes as input a list of attendees as well as the meeting location. 
At minimum an attendee is represented by a location (e.g. country), but can also be associated with other criteria such as organization, type of presence (e.g on-site, remote, ...). 
These criteria can be used to cluster attendees according to the different values of these criteria. 
Each value can be associated with an amount of \co{} equivalent emissions or the number of attendees.  
The \texttt{CityDB} class is responsible for associating an airport to a location.

The \co{} equivalent emissions of a flight is estimated by a \texttt{mode} ( i.e. 'distance' and 'flight') and \texttt{co2eq} the methodology ( i.e. 'myclimate2019'~\cite{myclimate} and 'goclimate2019'~\cite{goclimate}).
The 'distance' mode is solely based on the distance between the city of the meeting and the city of the attendee. 
The resulting \co{} equivalent emitted corresponds to a direct flight between these two cities -- thus ignoring detours, takeoff and landing operations associated with multi segment flights.   
The 'flight' mode, in return, considers a real flight between the two cities eventually with potentially multiple segments. 
The \texttt{FlightDB} class returns such flights by requesting the Amadeus 'Flight Offer Search' API~\cite{flightsearchoffer} that returns all available matching flights. 
The \texttt{AmadeusOffersSearchResponse} class is responsible for parsing that response and selecting a plausible flight.
The \texttt{Flight} class estimates the \co{} equivalent of the flight by considering each segment as an individual flight. 
\thiscoeq{} implements two methodologies to compute the \co{} equivalent, 'myclimate2019' and 'goclimate2019' -- as detailed in Section~\ref{sec:co2eq} . 
\texttt{Flight} directly implements the 'myclimate' methodology while 'goclimate' is implemented by requesting a GO Climate Neutral service. 

\begin{figure}
\centering
  \includegraphics[width=\columnwidth]{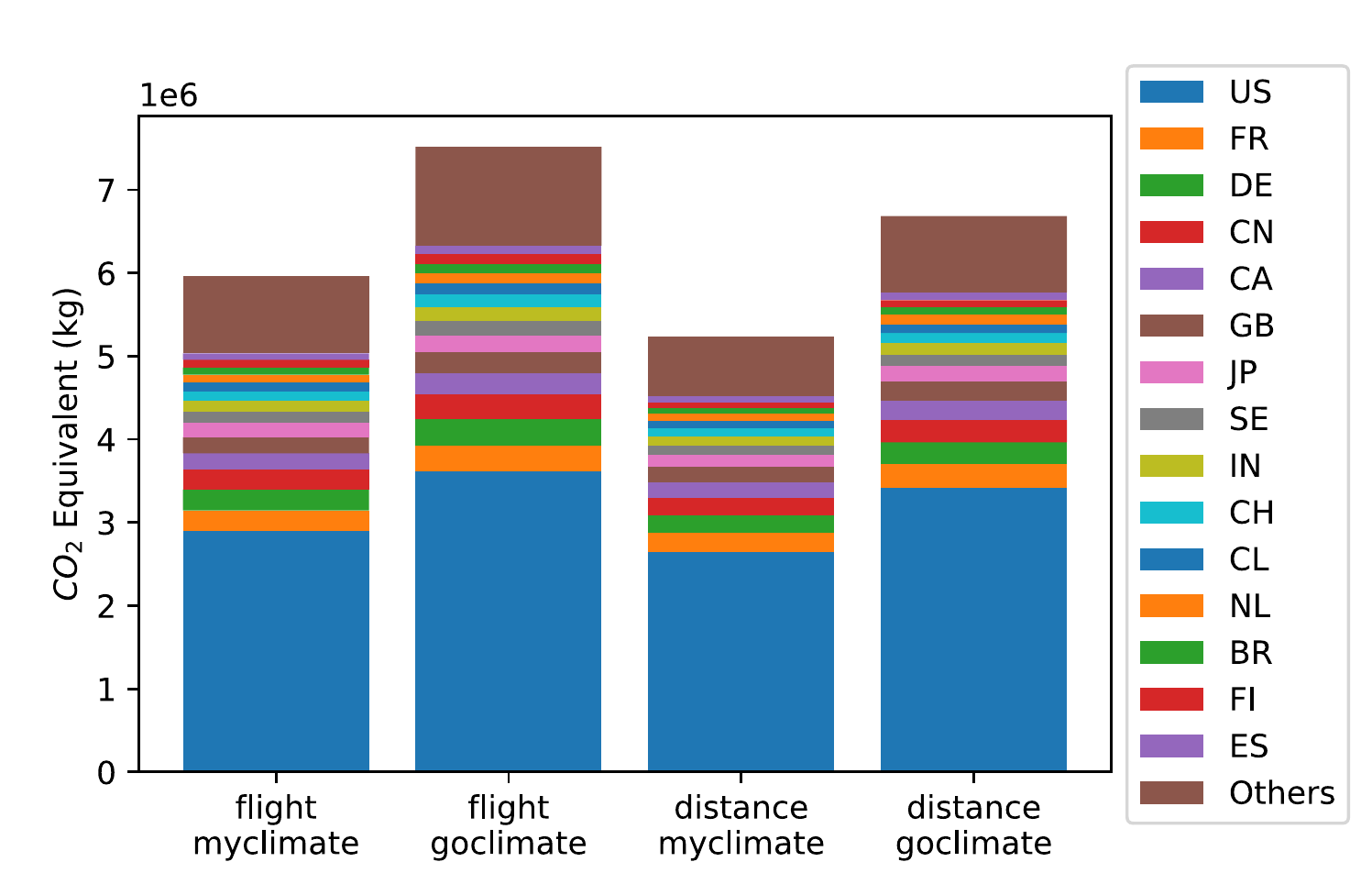}
\caption{Computation of \co{} equivalent emissions for IETF~100 with a representation of \co{} emissions clustered per country and estimated in kg. IETF~100 had a total number of 1618 attendees ('remote', 'not-arrived', 'on-site'). }
\label{fig:ietf100}
\end{figure}

In addition to the computation of \co{} for a single meeting, the class \texttt{MeetingList} visualizes the evolution of \co{} equivalent emissions across various meetings. 
The \texttt{IETFMeeting} and \texttt{IETFMeetingList} classes extend the \texttt{Meeting} and \texttt{MeetingList} classes, mostly to retrieve, parse and cleanup of the attendee list from the IETF web site.
An IETF attendee is represented as a dictionary with the following keys: 'organization', 'presence', 'country'. 
An additional element 'flight\_segments' that indicates the number of segments associated to flight is computed on the fly.
Attendees can be partitioned according to these keys, and for each possible value, it is possible to estimate the number of attendees or the \co{} equivalent emission. 
The \texttt{IETFMeetingList} class - as opposed to taking a list of Meeting objects - takes the list of all IETF meetings -- set as an global variable --, and instantiates IETFMeeting objects when these meetings have not been created.
In addition, it performs the necessary adjustment (size, title, labels, ...) to plot a relevant figure.
Figure~\ref{fig:ietf100} provides an example of the estimation provided by \coeq{} for a single meeting. 
For more examples of estimation provided for a list of meetings, please see section~\ref{sec:ietf} or ~\cite{co2eq.io} for an exhaustive and up to date list of \coeq{} outputs. 

\subsection{Design and Performance }

\begin{figure}[hbt]
\centering
  \includegraphics[width=\columnwidth]{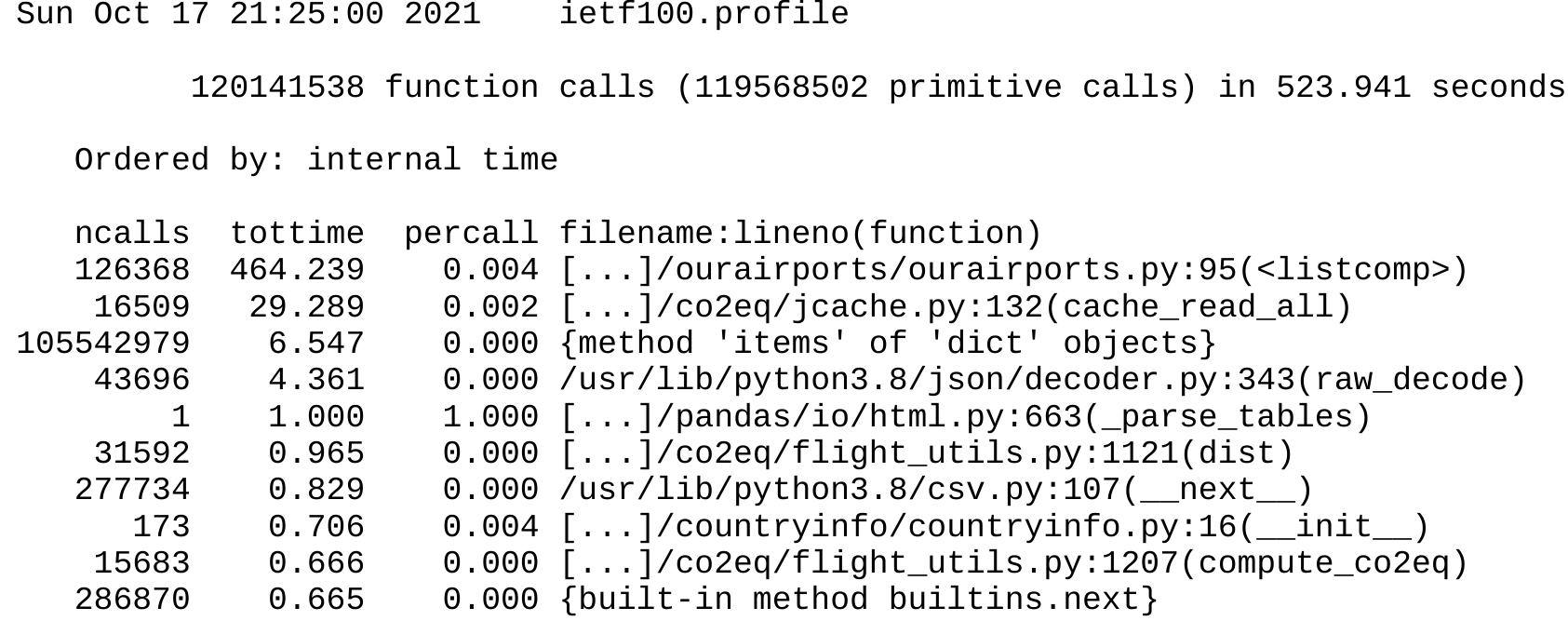}
\caption{Profiling the computation of IETF 100 with cProfiler}
\label{fig:cprofiler}
\end{figure}

\coeq{} is implemented in Python 3.8 as execution time is not especially crucial. 
We briefly evaluate the performances using cProfiler~\cite{cprofiler} as it does not require any changes to the code and estimate \co{} for the IETF~100 with all necessary information being cached.
As represented in Figure~\ref{fig:cprofiler}, the total computation takes 523.941 seconds with 464.239 seconds associated with the ourairports module and 35 seconds associated with the read\_all function of jcache module involved by the \texttt{CityBD} class. 
We suspect the \texttt{OurAirports} class from ourairports module performs search within a list and this for any airports of any segment. 
A \texttt{AirportDB} class should inherit from \texttt{OurAirports}  and implement dictionary search.  
Currently \texttt{CityDB} is still using a list of IATA cities, but we also expect this class to undergo some major function redesign -- see section~\ref{sec:evolution}.

We have not performed an extended analysis over \coeq{}, as performance is not the primary purpose. 
However, we have favored the use of dictionaries over lists to speedup search. 
The drawback is that list enables search using multiple search entries while dictionaries have a single entry key. 
This is especially true for flight offers that are retrieved using multiple parameters such as origin, destination, dates, classes.
In order to provide some sort of flexibility for the search, we used primary keys - in our case origin, destination - which refers to a list of possible keys to reduce the size of the list. 
We also limit the size of the cached objects, and only the latest resulting flight and input parameters are cached. 
In case the primary key matches but not the secondary parameters match the cached object a new search is performed. 
The search firstly looks whether a new flight can be derived from the list of offers stored in a file origin-destination.tar.gz. 
If the file cannot be found or the flight offers present in the file do not match the criteria, a new request is sent to the Amadeus 'Flight Offer Search' service.
The flight response is derived, cached and the additional offers are placed to the origin-destination.tar.gz.

\subsection{Evolution}
\label{sec:evolution}

Most foreseen evolution for \coeq{} is led by increasing the ability to 1) automatically and transparently handle various types of locations and usages as well as 2) to extend the \co{} evaluations. 

Estimation of \co{} requires the computation from a departure point and a destination point. 
In the 'distance' mode inter-city distance can be computed, but in the 'flight' departure and destination are airports. 
Both cities and airports are represented by IATA codes. 
The Amadeus Search Offer that takes IATA city code as input - as well as IATA airport and we use it to convert an IATA city code into the appropriate IATA airport code. 
While in many cases, IATA city code and IATA airport city code are the same, this is not always the case, as some large cities have multiple airports -- PAR for Paris is associated with multiple airports CDG, ORY.    
As a result, the main purpose of \coeq{} is to translate an attendee location into a city IATA code or an airport IATA code. 

In the case of the IETF, the attendee location is an ISO-3166 - alpha2 country code~\cite{iso-3166}.
There are currently 249 country codes that make it possible to assign -- even manually -- a country to an airport. 
The general strategy we adopted, is to derive the capital city name (a string) from the country and find the IATA city object with that capital name, and thus derive the associated IATA city code.
The binding of the ISO-3166 country code to the IATA city code results from a match between the capital name -- which is a string -- between two databases. 
One database that provides the capital from a country code and one database that contains the list of IATA cities.  

The match is possible if the country code is effectively considered as a country code -- by both CountryInfo~\cite{countryinfo} and ISO-3166~\cite{iso-3166} and if the capital name returned by CountryInfo corresponds to a name associated to a IATA city. 
We experienced issues with ISO-3166 that did not recognize the following country codes -- 'RS' (Serbia), 'ME' (Montenegro), 'MM' (Mayanmar) -- that were not referenced by the python module iso3166. 
We should probably update the module. 
In addition, some other countries may not have official capitals -- for example Palestine. 
For other countries, the capital is only administrative and does not represent well the hub of the nation which could result in flight search error.
For example, for Australia, we switch Canberra to Sydney, for the US we split the main cities randomly from WAS (Washington) and LAX (Los Angeles) and so on.  
For other countries such as Andorra or many territories, the main city is not within the country itself. In some cases, such as Andorra, the closest main city (Toulouse) is not even the capital of the other country (France). 
At last the name of the capital provided by CountryInfo does not match the one in the IATA cities due to different spelling or that the name is associated with the name of the territory instead of the capital.
Finally, in some cases, the airport provided was not able to offer flights, in which case we needed to approximate the location to another. 
Overall, the ISO-3611 country code to IATA city currently requires some manual adjustments and we would like to be able to provide a more robust approach especially as the approach would probably not scale to more diverse locations such as cities -- which is our intention. 
One foreseen alternative is to use geographic coordinates in combination of airport popularity or size in terms of passengers per year or a specific Amadeus service such as Airport \& City Search~\cite{airportandcitysearch}. 

On the one hand, we expect \coeq{} to continue to be extended to take every meeting's specificity, but we also expect \coeq{} to provide an easy way to be used by default. 
We are thinking of defining a common input JSON format for meeting attendees and meeting parameters -- especially for meeting lists -- to make the use of \coeq{} easy via a web interface. 

We would like to extend the \co{} estimation and include additional measurement - such as ICAO~\cite{icao} or updated models from myclimate and Go Climate.  
We also expect to complete \coeq{} by including \co{} emissions associated with hotels and meeting venues -- starting with~\cite{eia, epa, ghgp, ghgp-online}  -- as well as other transports. 
We would also like to be able to compute the \co{} associated with video conferences to better estimate the gains provided by remote meetings.  
At last we would like to extend \coeq{} to other usages than meetings which might be achieved by using a more generic data model -- at least internally as we do not want the specific case of estimation for meetings to reflect such complexity.

\section{Case Study: \co{} emission analysis for IETF meetings} 
\label{sec:ietf}
This section details \coeq{} outputs for the IETF meetings. Note that~\cite{co2eq.io} contains an exhaustive and up to date list of \coeq{} outputs. 
Section~\ref{sec:ietf-co2} is primarily focused on interpreting the environmental effect associated with handling on-site IETF meetings. 
\co{} equivalent emissions are estimated and confronted to the general perspective of the 2015 Paris agreement, the IPCC Working Group I contribution to the Sixth Assessment Report AR6-WG1~\cite{ar6-wg1} as well as general envisioned scenarios for aviation. 
Section~\ref{sec:ietf-segments} estimates the average number of flight segments per attendee and suggests that such data may lead to further investigations as to limit exposure to virus and limit the widespread of a pandemic. 
At last, Section~\ref{sec:ietf-others} depicts how \co{} may be used as a more generic metric to measure IETF growth and analyze some trends at the IETF such as diversity, transparency.

\subsection{\co{} Emissions and Climate Change}
\label{sec:ietf-co2}

This section estimates the amount of \co{} equivalent generated by IETF meetings over time and compares the average of \co{} equivalent emissions per participant to the average \co{} emissions per capita of various countries. 
Then, different scenarios that apply to  general aviation -- each associated with a specific increase of the global temperature -- are applied to the flights associated with IETF meetings using different meeting frequencies.  

\subsubsection{IETF attendee \co{} equivalent versus countries' \co{} emissions per capita}

\begin{figure}[hbt]
\centering
\begin{subfigure}{\columnwidth}
    \includegraphics[width=\columnwidth]{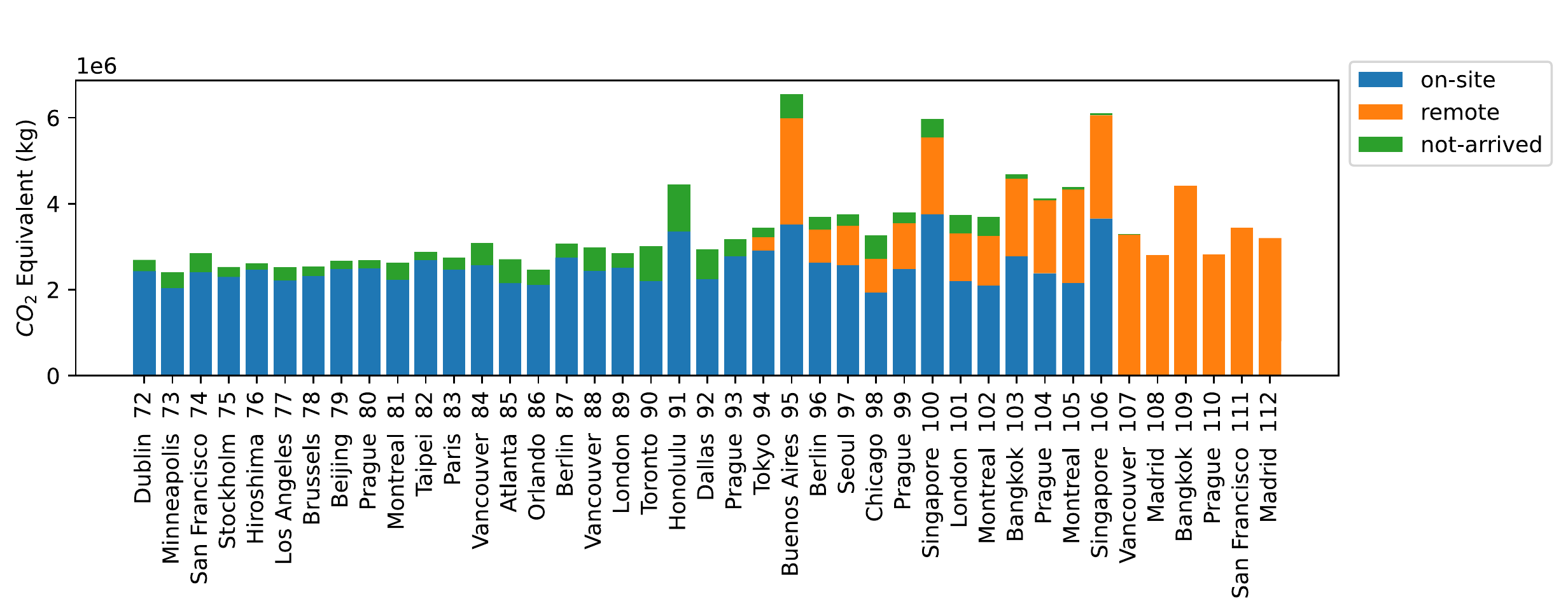} 
    \caption{'myclimate'}
    \label{fig:presence-myclimate}
\end{subfigure}
\\
\begin{subfigure}{\columnwidth}
    \includegraphics[width=\columnwidth]{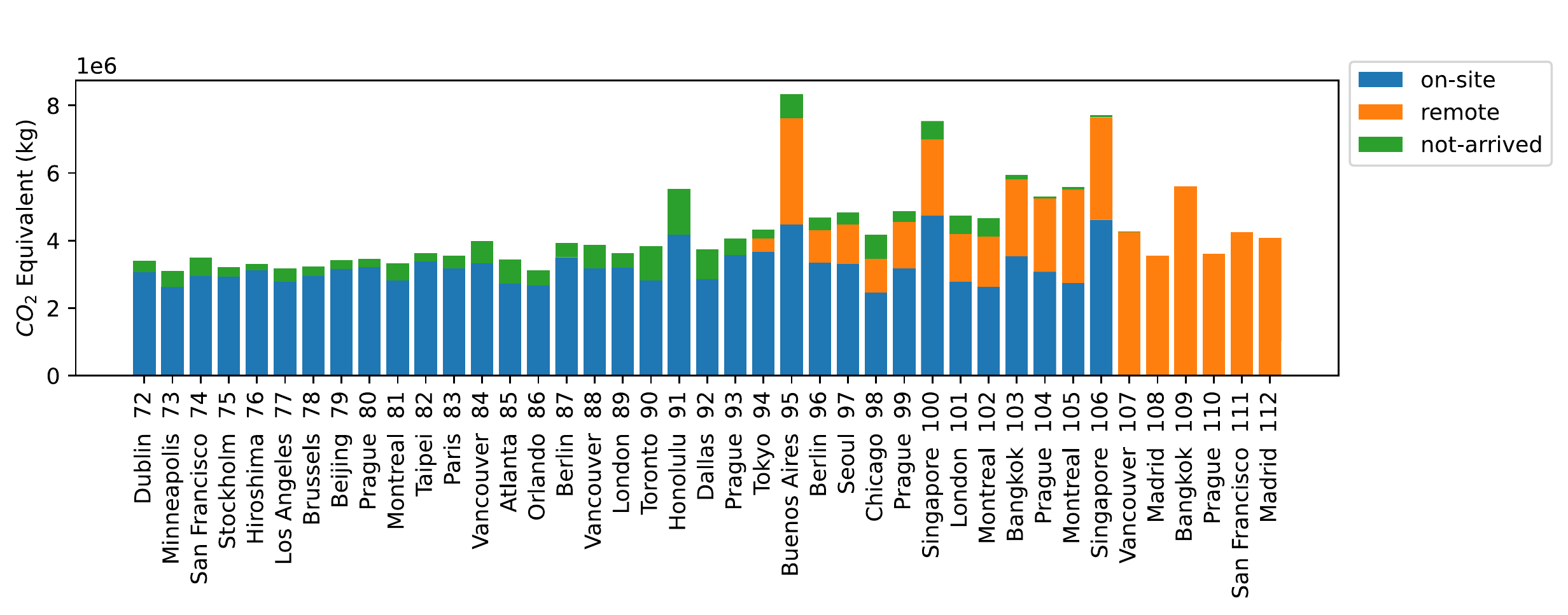}
    \caption{'goclimate'}
    \label{fig:presence-goclimate}
\end{subfigure}
\caption{Total \co{} emissions per presence type that is for 'on-site', 'remote' and 'not-arrived' attendees}
\label{fig:presence}
\end{figure}

Figure~\ref{fig:presence} depicts the evolution since IETF~72 in Dublin of \co{} emissions equivalent associated to air traffic based on estimated -- but real -- flight itineraries. 
For each flight the \co{} equivalent is estimated according to the 'myclimate' and 'goclimate' methodologies --  both described and compared in section~\ref{sec:co2eq}. 
Attendees are then clustered according to their type of presence ('on-site', 'remote' or 'not-arrived'). 
While all attendees are being assigned air flight, the effective \co{} emissions of the meeting are represented by 'on-site' participants only. 

The effective amount of \co{} equivalent emissions for IETF meetings are quite stable between 2.5 and 3 Gg from IETF~72 in Dublin to IETF~93 in Prague. 
During this period, meetings in North America tend to provide a slightly lower amount of emissions -- but not always and IETF~91 in Honolulu is an outlier with significantly more emissions.
From IETF~94 in Tokyo to IETF~106 in Vancouver, the amount of \co{} equivalent emissions presents a slight decrease with peaks associated to Asian locations. 
IETF~107 Vancouver to IETF~112 in Madrid were entirely virtual with no 'on-site' participation. 

The average effective \co{} equivalent emissions from  IETF~72 to IETF~106 is estimated to be 2.5 Gg by myclimate and 3.2 Gg with goclimate which corresponds respectively to an average of 2.2 and 2.7 tonnes per attendee.

\begin{figure*}
\centering
\begin{subfigure}{0.5\textwidth}
  \includegraphics[width=\columnwidth]{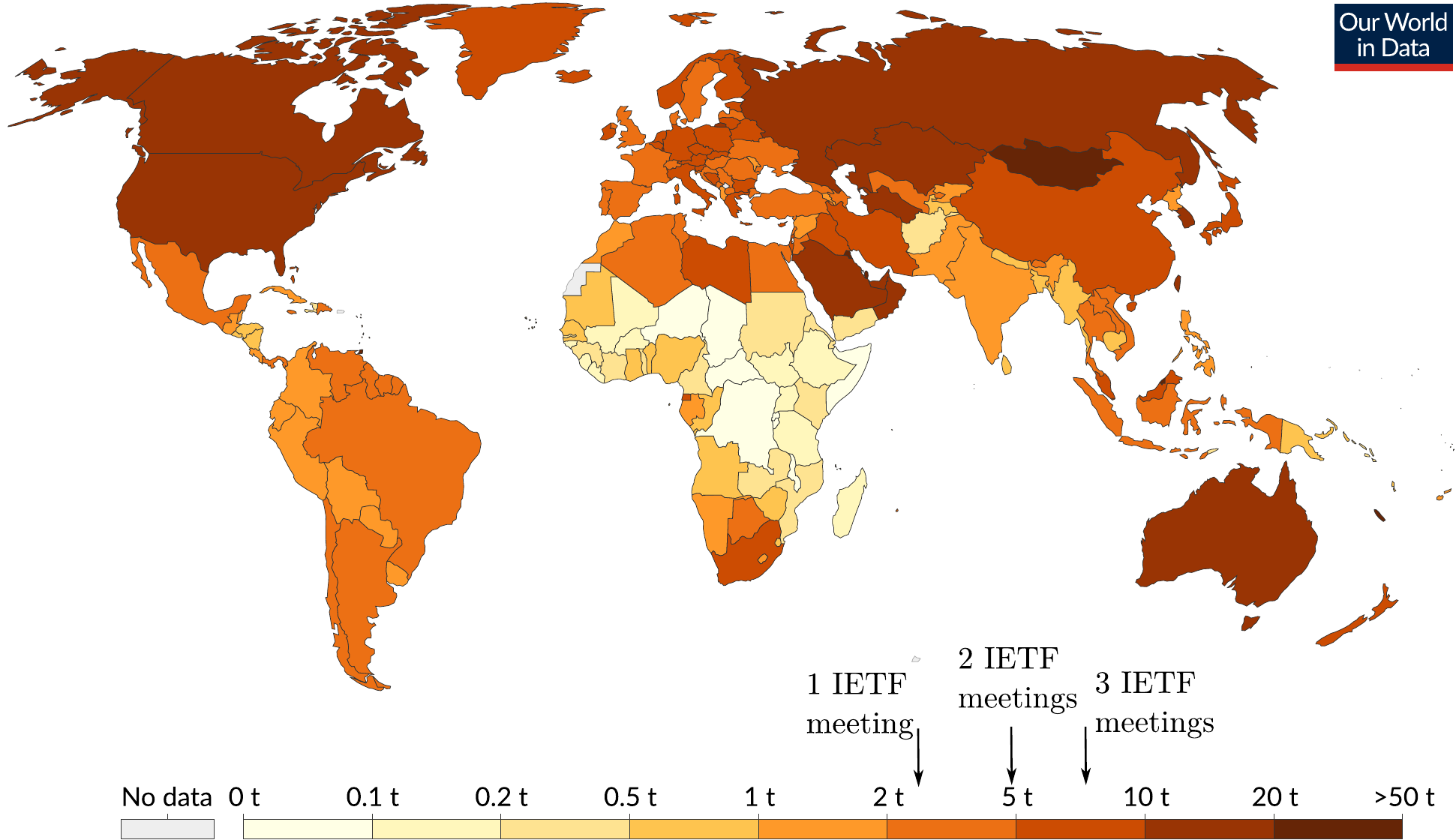}
  \caption{World Map view}
  \label{fig:percapita-map}
\end{subfigure}
\\
\begin{subfigure}{\textwidth}
\centering
  \includegraphics[width=\textwidth]{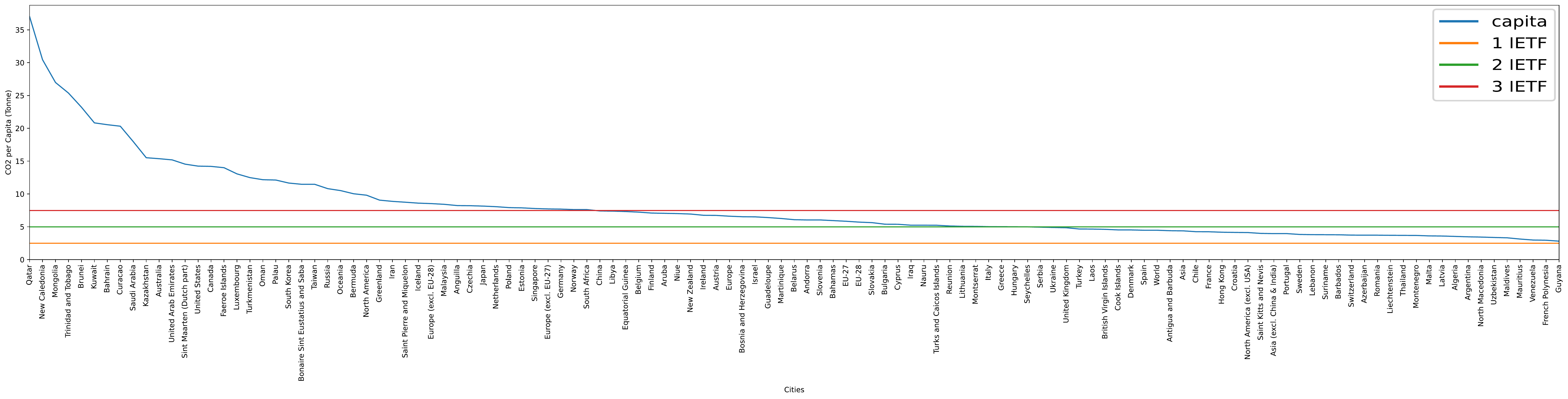}
\caption{ \co{} emissions of IETF participants in regard with \co{} emissions per capita -- Representing 116 countries out of 229.}
  \label{fig:percapita-chart}
\end{subfigure}
\caption{ \co{} emissions are evaluated from the burning of fossil fuels for energy and cement production. Land use change is not included. \co{} are measured on a production basis, meaning they do not adjust for emissions embedded in traded goods. Data and world Map are provided by Our World in Data based on the Global Carbon Project~\cite{owidco2andothergreenhousegasemissions} }
\label{fig:percapita}
\end{figure*}

Figure~\ref{fig:percapita-map}~\ref{fig:percapita-chart} compares attending 1, 2 and 3 IETF meetings a year to the annual \co{} emissions per capita provided by~\cite{owidco2andothergreenhousegasemissions}~\cite{gcp-2021}~\cite{essd-2021-386}.  
The \co{} equivalent emissions associated with the attendance of 3 IETF meetings a year corresponds to the emissions per capita of Germany, Finland, Poland, Belgium. 
These countries provide a higher amount of \co{} than the average European countries mostly due to the use of coal to generate their energy.
The \co{} equivalent emissions associated with the attendance of 2 IETF meetings a year corresponds to the per capita emissions of European countries such as Greece, Italy, UK. 
The \co{} equivalent emissions associated with the attendance of a single IETF per year corresponds to the emissions per capita of countries such as  Venezuela, Mauritius.   

\subsubsection{Comparing IETF air flight traffic with envisioned scenario for aviation}

\cite{aviation-2021} suggests aviation will contribute to 0.1 C of warming in 2050 if pre COVID-19 aviation growth would resume. 
It further analyses 4 types of scenarios to foresee the future of aviation with their respective responsibility and contribution to the increase of temperature in 2050.
These scenarios assume a post-COVID-19 recovery growth until 2024 followed by a post 2024 growth. 
The 'no pandemic' scenario considers no pandemic occurred  with air travel growing by 3\% per year since 1970. 
This scenario results in aviation being responsible for raising temperature by 0.1~\degree C.
The 'back to normal scenario' considers a post COVID-19 growth of 16\% per year and 3\% per year thereafter. 
This scenario results in aviation being responsible for raising temperature by 0.09~\degree C in 2050. 
More importantly it shows that the brutal and forced decrease of flights during the COVID-19 has very little long term impact. 
The 'zero long term growth' assumes a post COVID-19 recovery growth of 13\% followed by a 0\% growth. 
This scenario is responsible for raising temperature by 0.06~\degree.
Finally the 'long term decline' assumes a post COVID-19 recovery growth of 10\%  followed by a -2.5\% growth, which ends up in air traffic level decreased by 50\% compared to 2019 -- that is the level during the pandemic. 

\begin{figure}
\centering
    \includegraphics[width=\columnwidth]{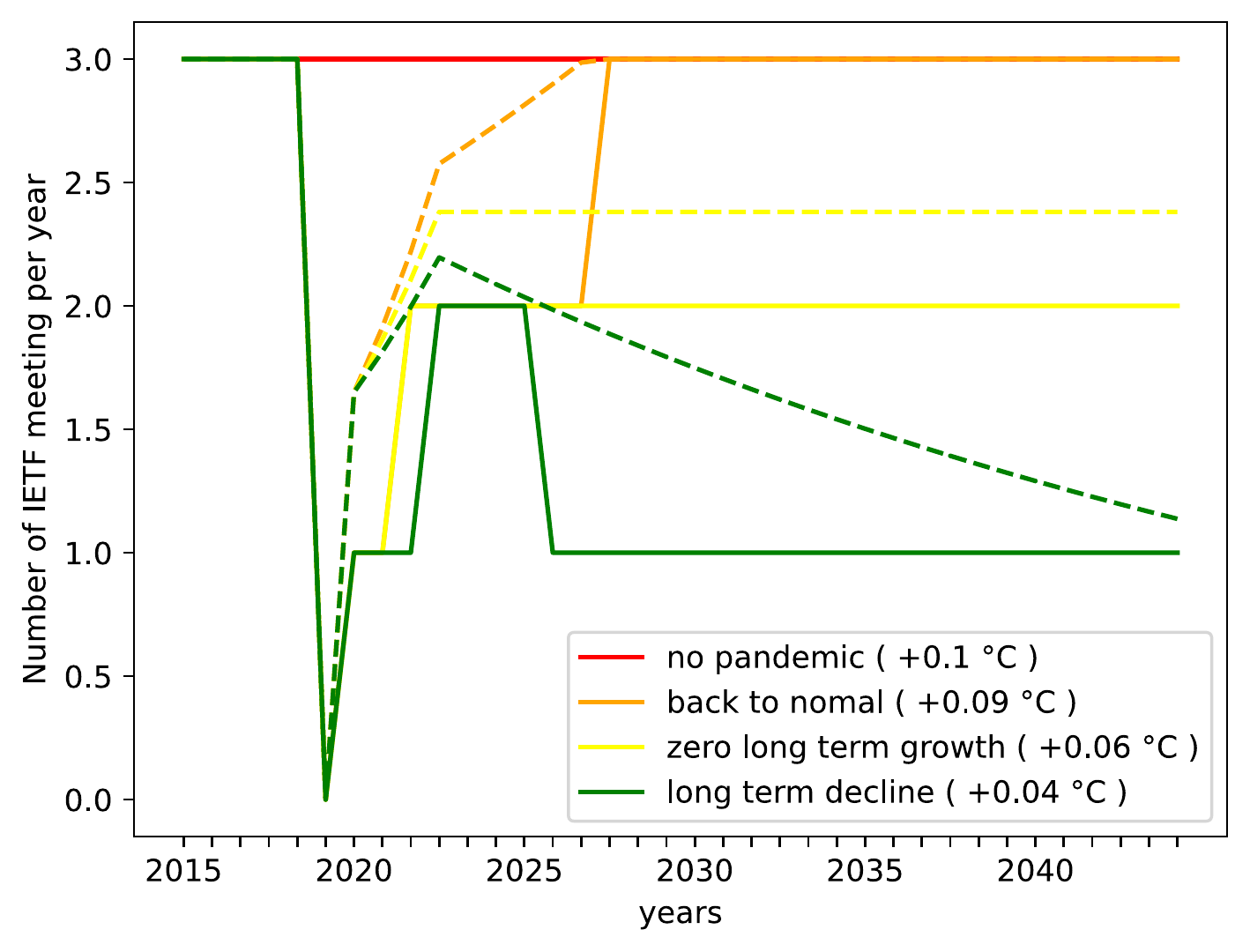}
\caption{Application of the scenarios described by~\cite{aviation-2021} to IETF meetings}
\label{fig:aviation2021}
\end{figure}

While these scenarios apply for the whole air traffic, Figure~\ref{fig:aviation2021} apply these scenario to the IETF meetings, assuming the same number of participants during the meetings and considering for 2021 only a 45\% decrease over 2019 -- as opposed to a 100\% decrease that has been observed with IETF meetings being fully virtual. 
Another adaptation is that unlike aviation growth the number of meetings is not expected to be greater than 3 meetings per year. 
The dash lines show fractions of meetings which may be useful for further studies considering hybrid meetings, that is when a significant fraction of the attendees are 'remote'. 
However, this is left for further analysis.
Application of the air traffic scenarios to the IETF related air traffic shows that scenario ignoring the needed effort to fight climate change ('no pandemic' and 'back to normal') results in 3 meetings a year for the IETF while other scenarios ('zero long term growth' and 'long term decline') results in respectively 2 or 1 IETF meeting a year. 

In 2015 nations agreed to limit global warming well below 2~\degree C. 
Current forecast based on Nationally Determined Contributions (NDC) established that we are heading toward 2.4~\degree C. 
The IPCC Working Group I contribution to the Sixth Assessment Report AR6-WG1~\cite{ar6-wg1} insisted that every fraction of a degree of increase is important and that major effort needs to be done to reach the achievable 1.5~\degree C.
In such a context, it seems inappropriate to maintain a rate of 3 IETF meetings a year by which participants produce as much emissions as European countries using coal to generate their energy.  
A more sustainable approach is needed and the target to reduce by 45\% emissions in the Paris agreement in 2015 as well as a sustainable scenario for aviation suggest  limiting IETF meetings to 1 IETF meeting a year as well as huge effort to improve the remote participation experience. 

\subsection{Limiting Air Flights Connections to Limit Virus Exposure}
\label{sec:ietf-segments}

While the COVID-19 pandemic situation took us by surprise, it is likely that pandemic frequency increases and that more severe pandemics are to come - especially as the root cause of pandemic is due to the anthropogenic destruction of biodiversity~\cite{ipbes-2020}.  

\begin{figure}[htb]
\centering
    \includegraphics[width=\columnwidth]{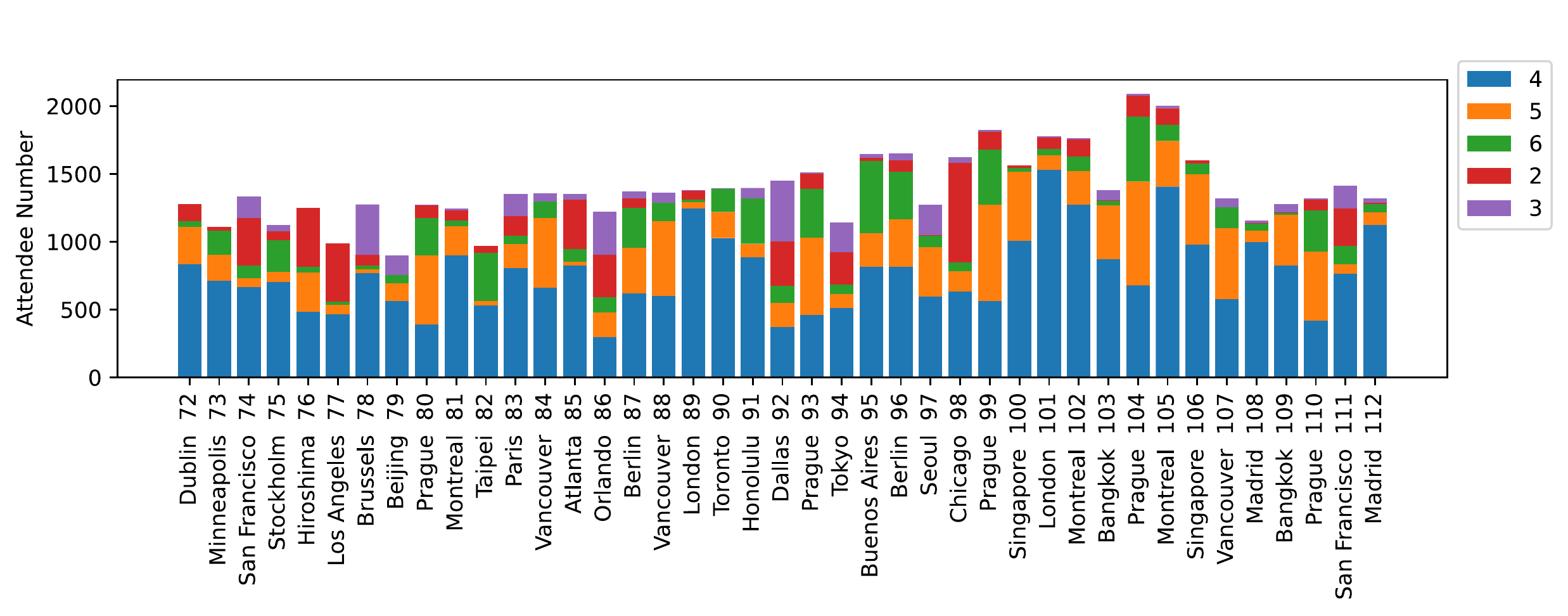}
\caption{Grouping participants per the number of their associated flight segments number. }
\label{fig:segments}
\end{figure}

While work remains to be done to evaluate the exact role airports are playing in the spreading of a pandemic, it remains plausible that limiting the number of transit airports reduces the risk of infection and consequently the widespread of the pandemic.
Figure~\ref{fig:segments} depicts for each IETF the number of flight segments for each attendee -- including 'on-site', 'remote' and 'not-arrived' --- and Table~\ref{tab:segments} orders the IETF meetings according to the average number of segments per attendee.
It appears that places like Tokyo, Los Angeles and Bejin are the destinations that minimize flight connections.

Of course, such findings require additional analysis to refine, for example, the role of airports into the widespread of a pandemic, refining the total number of connections to the number of connections in international airports, the duration of the connection, the time to retrieve luggage among other things.
It should also consider that departure location has been estimated as the capital of the attendee's country which may for large countries introduce a bias. 

\begin{table}[htb]
\begin{tabular}{|p{2.3cm}llp{1.5cm}|}
\hline
IETF Meeting & City       & Country & Connections \\
\hline
77 & Los Angeles & US & 2.4 \\
94 & Tokyo & JP & 2.8 \\
79 & Beijing & CN & 2.8 \\
98 & Chicago & US & 3.2 \\
83 & Paris & FR & 3.4 \\
89, 101 & London & GB & 3.5 \\
92 & Dallas/Fort W & US & 3.5 \\
74 & San Francisco & US & 3.5 \\
86 & Orlando & US & 3.5 \\
85 & Atlanta & US & 3.6 \\
78 & Brussels & BE & 3.6 \\
76 & Osaka & JP & 3.6 \\
111 & San Francisco & US & 3.7 \\
97 & Seoul & KR & 3.8 \\
96, 87& Berlin & DE & 3.9 \\
75 & Stockholm & SE & 3.9 \\
72 & Dublin & IE & 4 \\
90 & Toronto & CA & 4.0 \\
108, 112& Madrid & ES & 4.1 \\
103, 109 & Bangkok & TH & 4.1, 4.3 \\
 81, 102, 105 & Montreal & CA & 4.1 \\
100, 106 & Singapore & SG & 4.2, 4.3 \\
95 & Buenos Aires & AR & 4.3 \\
82 & Taipei & TW & 4.3 \\
73 & Minneapolis & US & 4.4 \\
84, 88, 107 & Vancouver & CA & 4.4, 4.5, 4.6 \\
91 & Honolulu & US & 4.5 \\
80, 93, 99, 104, 110 & Prague & CZ & 4.5, 4.6, 4.7 \\
\hline
\end{tabular}
\caption{Ordered average flight connections per attend for each IETF meetings }
\label{tab:segments}
\end{table}

\subsection{Measuring Growth, Diversity and Transparency}
\label{sec:ietf-others}

\co{} equivalent emissions of an attendee can be seen as a metric that measures participation by attributing a cost to a given participation.   
The cost in question is obviously an environmental cost, but it also combines travel distance, travel expenses (air flight and hotel) as well as other costs such as time commitment to attend the IETF meeting.
Overall the sum of the attendee costs may reflect the value of the meeting and by extension the value associated by a certain type of participation. 
More precisely, the global cost associated with the participants could reflect the worth of an IETF meeting, and the cost associated with 'on-site' participation (respectively  'remote', 'not-arrived')  reflects the worth -- and share -- associated with each type of participation. 
The \co{} metric is very similar to the number of attendee number metric, however, the attendee number reflects an attendee decision while the cost estimation ponders attendee with the cost. 
More specifically, it provides more weight to distant participants, for which the participation has a higher cost. 
It may also provide a zero cost to attendees of the country when the IETF meeting is hosted in the country's capital.   
Overall this seems to lower attendees' participation due to their local presence at an IETF meeting.
Note that we are not trying to defend the \co{} metric as opposed to  the number of participants. 
Instead we are considering this metric as possibly providing a new angle that may be interesting. 

\begin{figure}
\centering
    \includegraphics[width=\columnwidth]{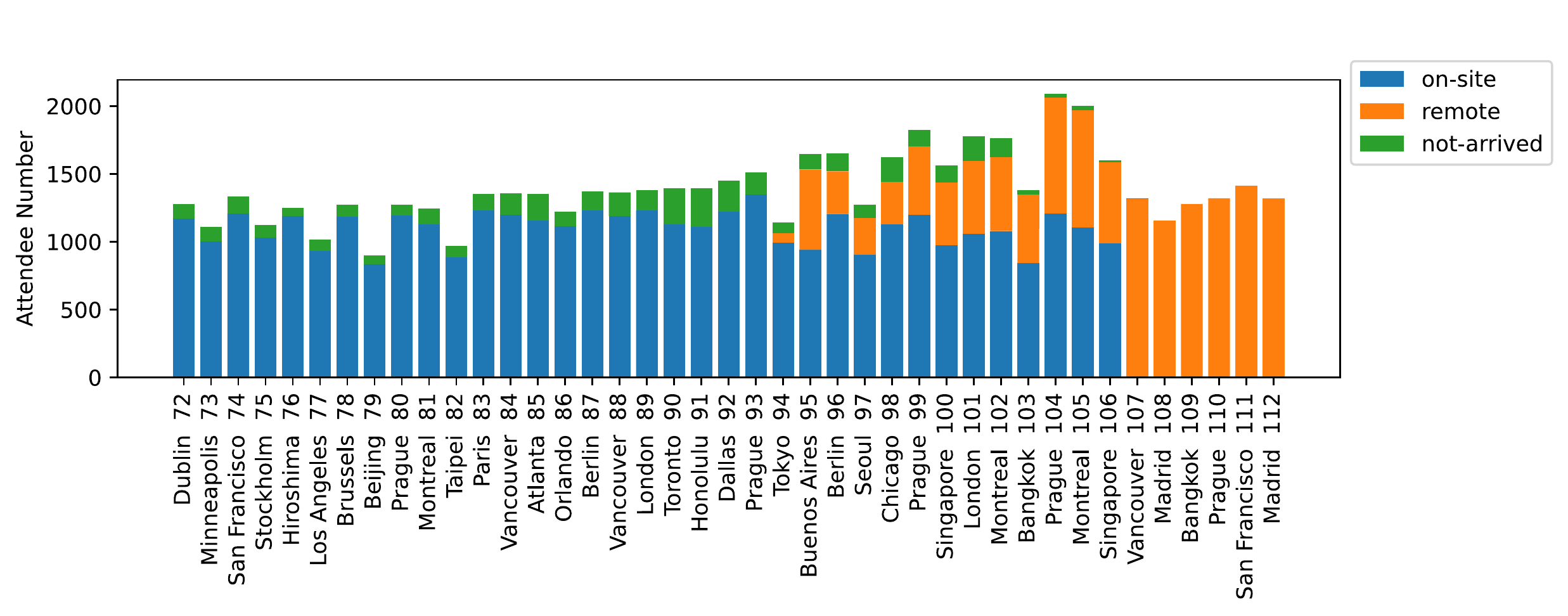}
\caption{Number of attendees clustered by presence type: 'on-site', 'remote' and 'not-arrived'. See equivalent figure with \co{} in Figure~\ref{fig:presence} }
\label{fig:presence-person}
\end{figure}

Figure~\ref{fig:presence} as well as Figure~\ref{fig:presence-person} respectively depict the \co{} emissions and the number of attendees per type of meeting participation. 
All three figures tend to show a similar trend that is: 'on-site' attendance is slowly declining and 'remote' participation is growing. However such trends are more visible using the \co{} metric as opposed to the number of attendee metric. 
Asian and South American (Buenos Aires) locations present peaks with the \co{} metric for both 'on-site' and 'remote' attendees while such peaks are less evident with the number of attendees metric. 
One possible way to interpret these peaks is that attendees are heavily located in North America and Europe. 

\begin{figure}
\centering
\begin{subfigure}{\columnwidth}
    \includegraphics[width=\columnwidth]{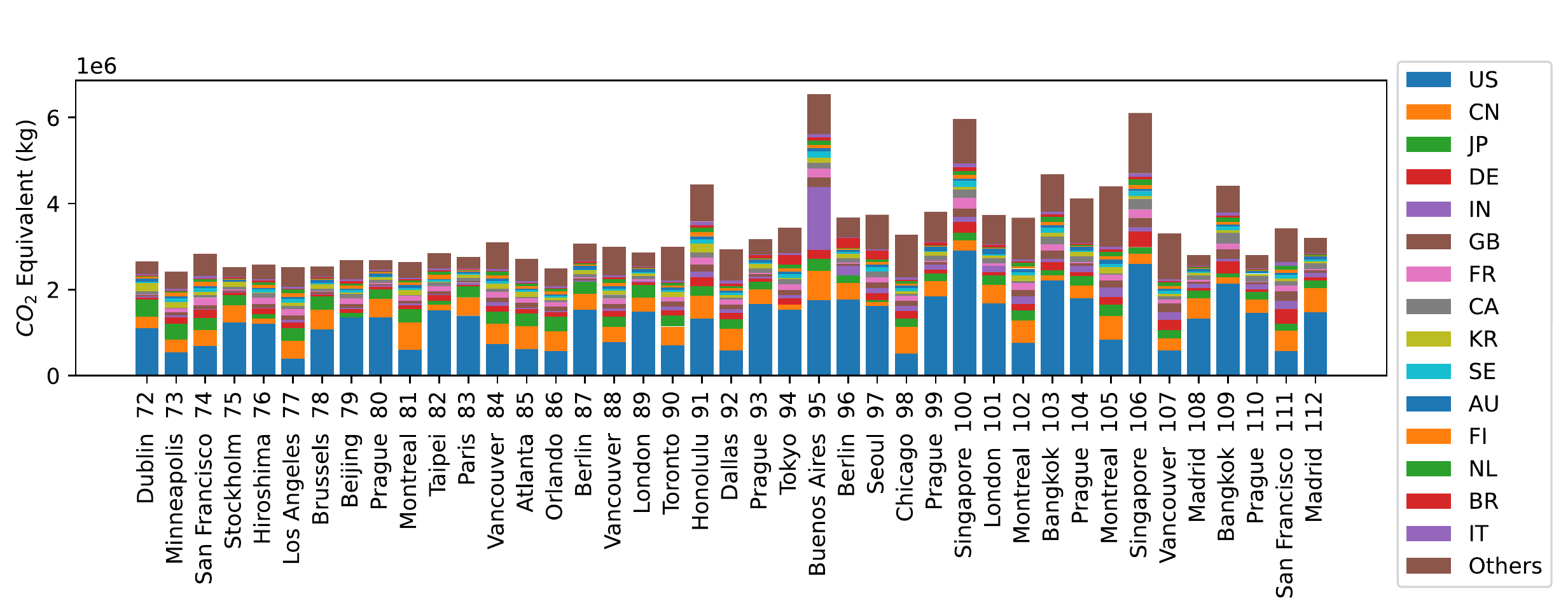} 
    \caption{\co{} estimated with 'myclimate'}
    \label{fig:country-myclimate}
\end{subfigure}
\\
\begin{subfigure}{\columnwidth}
    \includegraphics[width=\columnwidth]{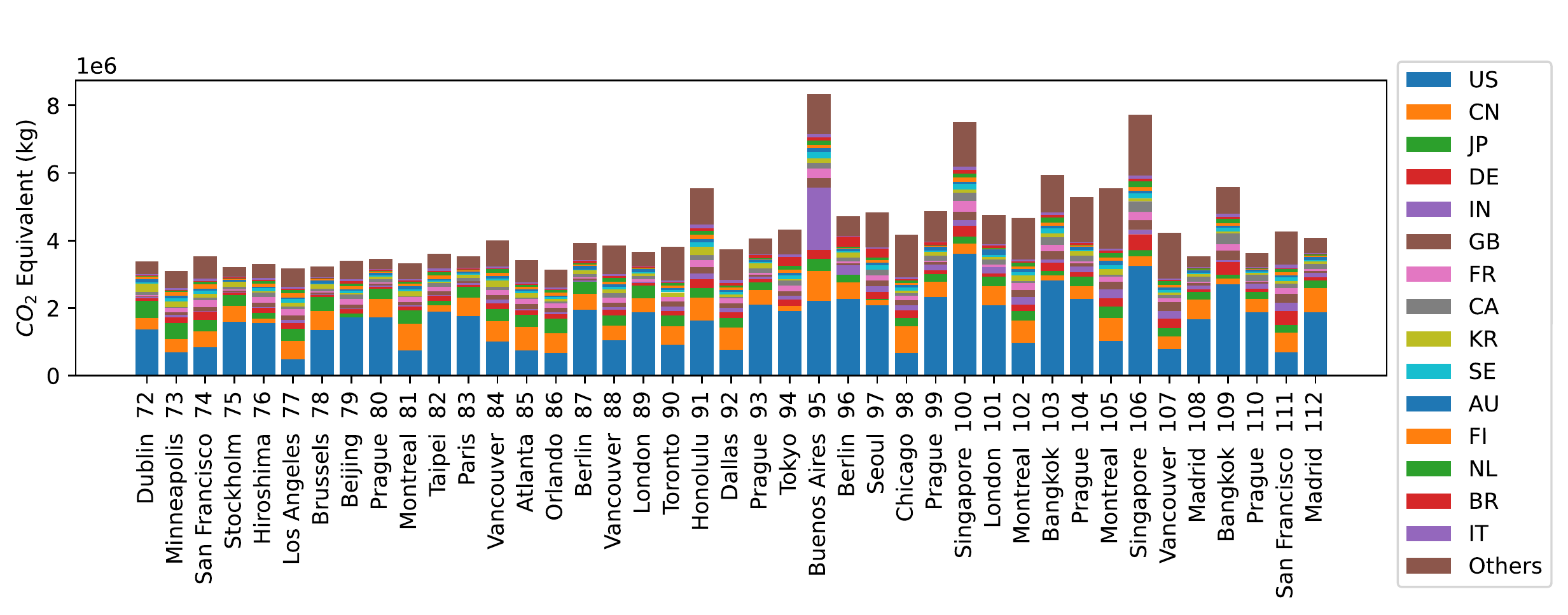}
    \caption{\co{} estimated with 'goclimate'}
    \label{fig:country-goclimate}
\end{subfigure}
\\
\begin{subfigure}{\columnwidth}
    \includegraphics[width=\columnwidth]{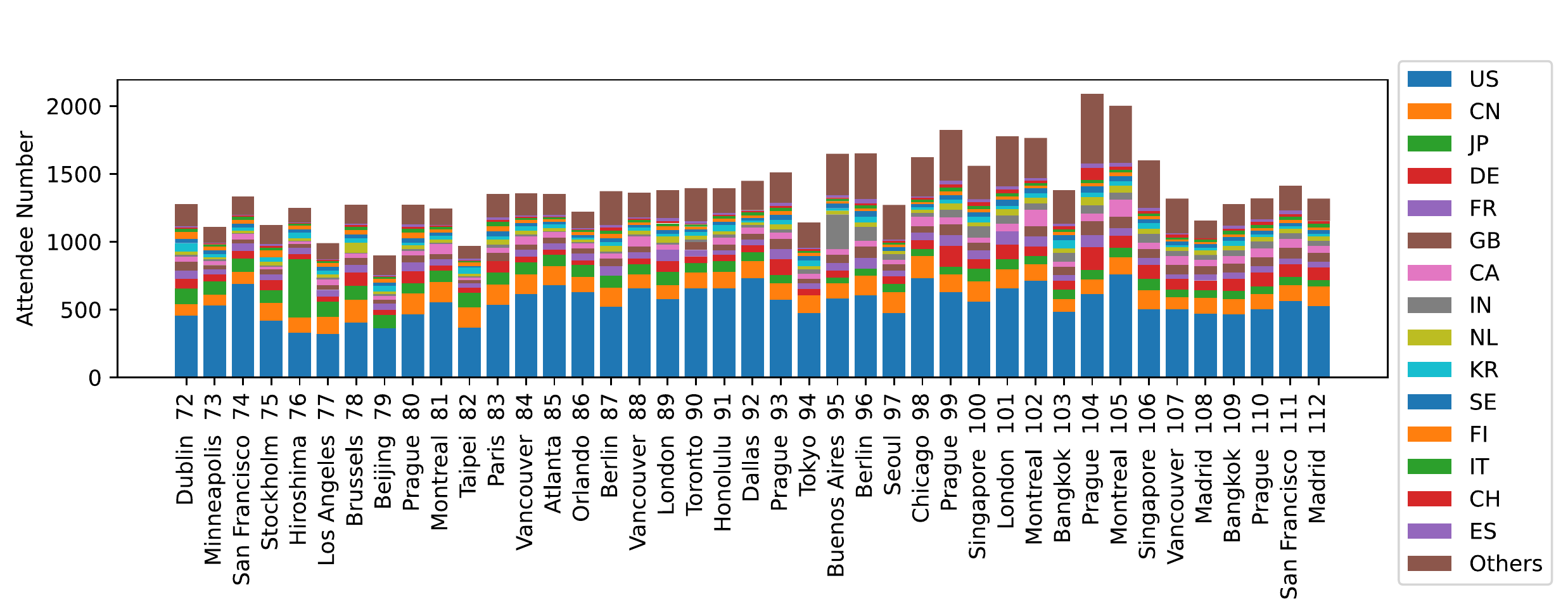}
    \caption{number of attendees}
    \label{fig:country-attendee}
\end{subfigure}

\caption{\co{} emissions and number of attendees clustered by countries. Only the 15 most represented countries are represented. }
\label{fig:country}
\end{figure}

Figure~\ref{fig:country} clusters \co{} emissions as well as the number of attendees per country.   
Both metrics show a large representation from the US compared to the other countries. 
On the other hand, Asia is well represented with China, Japan and Korea being the second, third and tenth most represented countries and overall the Asian region seems to be represented similarly to Europe. 
As countries present a huge difference in terms of population a representation in terms of region might be useful. 
However, from the country representation, it can be inferred that the African, the Middle East and the South American regions are under-represented. 
Figure~\ref{fig:country} also shows that the overwhelming majority of the attendees are mostly representing 15 countries, which seems to indicate a reaching out strategy may not be limited to regions, but may consider a finer granularity such as countries.   

\begin{figure}
\centering
\begin{subfigure}{\columnwidth}
    \includegraphics[width=\columnwidth]{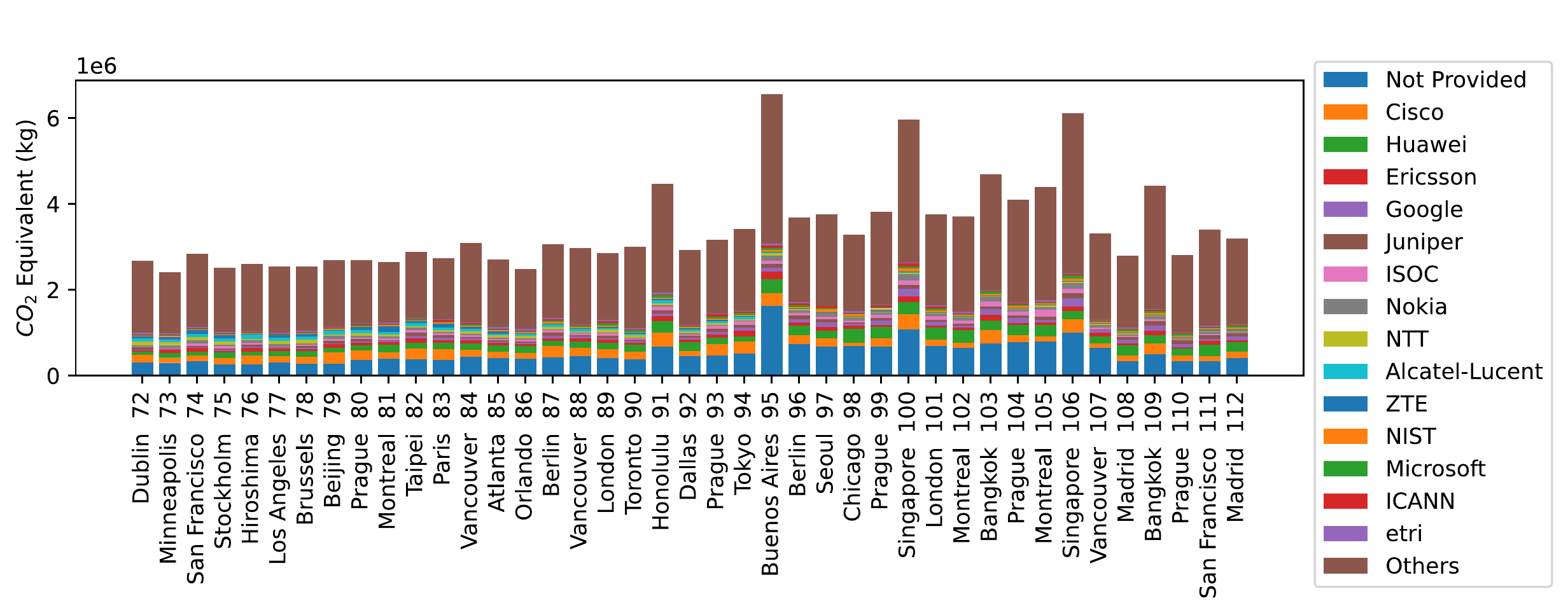} 
    \caption{\co{} estimated with 'myclimate'}
    \label{fig:organization-myclimate}
\end{subfigure}
\\
\begin{subfigure}{\columnwidth}
    \includegraphics[width=\columnwidth]{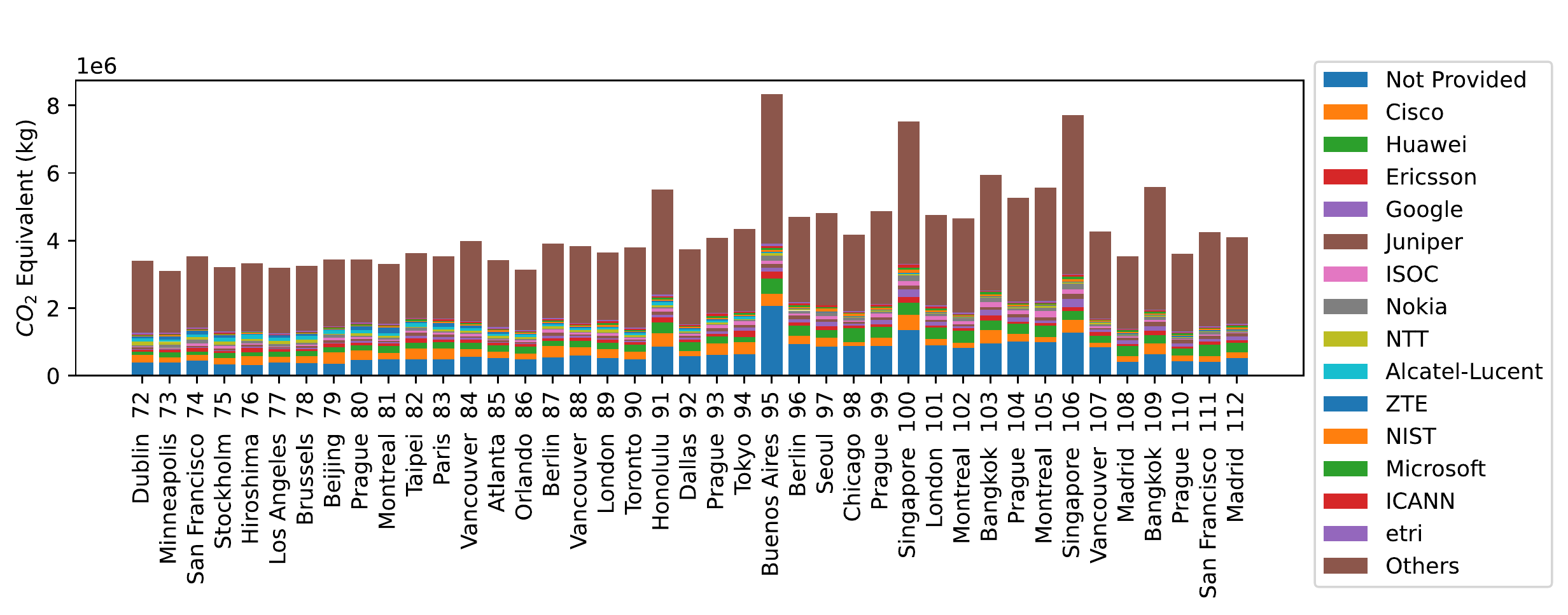}
    \caption{\co{} estimated with 'goclimate'}
    \label{fig:organization-goclimate}
\end{subfigure}
\\
\begin{subfigure}{\columnwidth}
    \includegraphics[width=\columnwidth]{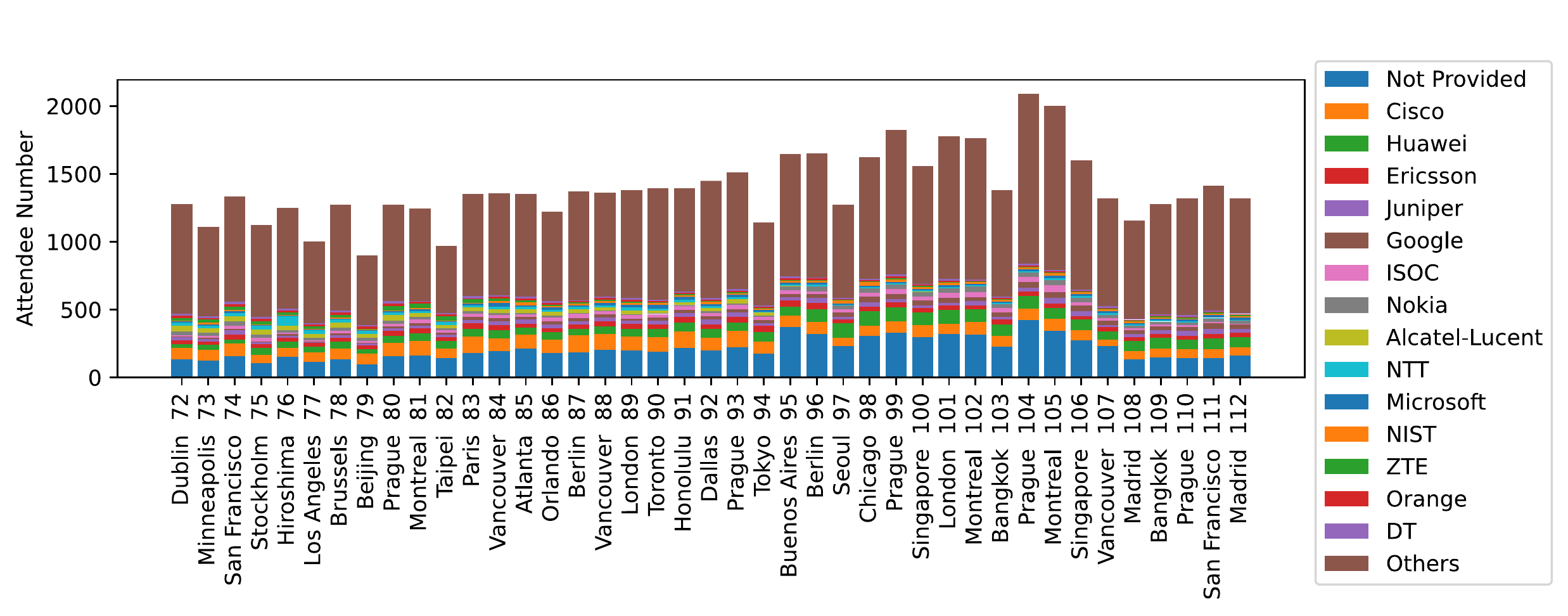}
    \caption{number of attendees}
    \label{fig:organization-attendee}
\end{subfigure}
\caption{\co{} emissions and number of attendees clustered by organization. Only the 15 most represented companies are represented. }
\label{fig:organization}
\end{figure}

Figure~\ref{fig:organization} clusters the costs and attendee numbers per organization.
The positive aspect is that the most represented organizations do represent less than 50\% of the global attendance.
On the other hand, further investigations are needed to understand the full ecosystem, that is whether organizations labeled as 'Others' are independent organizations as opposed to working for other declared organizations. 
Similarly, the most represented organization is the one labeled 'Not Provided' which indicates the field organization has not been filled by the attendee. 
Further investigations are also needed here to clarify the reasons this field is omitted. 

\section{Conclusions}
\label{sec:concl}
This paper presents \coeq{} that estimates the \co{} equivalent emissions generated by attending international meetings. 
Currently \thiscoeq{} limits its evaluation to air traffic emissions -- which is known to be the largest source of emissions for such meetings. 
\thiscoeq{} is an early version and many directions to make the tool evolve have been considered. 

IETF conferences have been the first case to be tested by \coeq{}.
The IPCC Working Group I contribution to the Sixth Assessment Report AR6-WG1 urged every sector to reduce its \co{} emissions to keep the target of 1.5~\degree C. 
While we are currently on track to a 2.4~\degree C increase in temperature, the 1.5~\degree C remains possible to reach. 
\coeq{} helps to estimate the IETF contribution and what efforts could be considered to remain a responsible community. 

The amount of \co{} emissions per attendee for 3 'on-site' IETF meetings per year corresponds to the average \co{} per capita emitted by European countries producing energy based on coal such as Germany or Poland.  
The amount for a 2 meeting attendance corresponds to the average \co{} per capita emitted by European countries such as UK, Greece or Italy, and the attendance to a single IETF meeting per year corresponds to the \co{} per capita of countries like Venezuela  or Mauritius. 
It is unsure there are substantial justifications for the IETF to contribute to that extent to the world wide \co{} emissions and  limiting  IETF 'on-site' meeting to at most 1 per year should be considered.  

Such a limit of 1 meeting per year is also the limit provided when considering the Paris agreement that required a 45\% decrease, as well as more recent studies evaluating the evolution of aviation in the next coming years. 
More specifically, 3 IETF meetings a year matches the evolution to the air traffic that would result in increasing the temperature between 0.09~\degree C and 0.1~\degree C.  
2 IETF (resp. 1 IETF) meetings a year matches scenarios where aviation would be responsible to increase the temperature of 0.06~\degree C (resp. 0.04~\degree C).

The IETF has already demonstrated during the COVID pandemic it can operate with only remote meetings. 
The main drawback that has been raised regarding 'remote' meeting is that hallway discussions are harder to happen. 
This is correct with the current IETF setting where most people attend the virtual meeting, by attending a specific session and then disconnect themselves when the session is done. 
The IETF has set a dedicated place for hallway discussion -- namely Gather -- but the current setting requires a specific connection to such a place which is definitely a different concept than a hallway where attendees have to go through to attend their sessions. 
Time zone is also an issue, and the IETF leadership has held many (successful) experiments in terms of agenda scheduling. 
Moving an organization from entirely 'on-site' to entirely 'remote' will take some time and while the feasibility is no longer to prove, some work remains to improve the remote experience. 
It also appears that 'remote' participation is a key factor to make the IETF grow and as such, seems a promising path.

That experimentation of 'remote' meetings -- while entirely feasible -- came so late and as such highlights that the IETF needs to increase its efforts on corporate sustainability.
Corporate sustainability would ensure the IETF operates in ways that, at a minimum, meet fundamental responsibilities. 
For example, the United Nations Global Compact~\cite{ungc} set 10 principles to address corporate sustainability in a broad sense, and only principles 7-9 are related to the environment. 
Adherence to such a program ensures these principles are part of the IETF strategy and becomes part of the IETF culture with the publication of Communication on Progress. 
More specifically related to \co{} emissions, the IETF may also consider adhering to the caring for climate initiative~\cite{c4c} led by the Global Compact, UN Environment Program (UNEP) and the secretariat of the UN Framework Convention on Climate Change (UNFCCC).


\section*{Acknowledgments}
We would like to thank Marie-Jose Montpetit for her feedbacks and the suggestion to consider the number of flight connections as a potential mean to provide safer travel. I also would like to thank Makan Pourzandi and Pernilla Bergmark for supporting and providing future directions. 

\ifthenelse{\boolean{acm}}{
  \bibliographystyle{ACM-Reference-Format}
}{
  \bibliographystyle{IEEEtranS}
}
\bibliography{bib}
%
%

\end{document}